\documentclass[12pt,a4paper]{article}
\usepackage{ucs}
\usepackage[utf8x]{inputenc}
\usepackage[english]{babel}
\usepackage[T1]{fontenc}
\usepackage{ae,aecompl}
\usepackage[usenames,dvipsnames]{color}
\usepackage[active]{srcltx}
\usepackage[numbers,comma,square,compress]{natbib}
\usepackage{amsmath,amssymb,amsfonts}
\usepackage{braket}
\usepackage{ifpdf}
\ifpdf
  \usepackage[pdftex,a4paper,hmargin=25mm,vmargin=24mm]{geometry}
  \usepackage[pdftex]{hyperref}
\else
  \usepackage[dvips,a4paper,hmargin=25mm,vmargin=24mm]{geometry}
  \usepackage[dvips]{hyperref}
\fi

\newcommand{\email}[1]{\href{mailto:#1}{#1}}

\newcommand{\be}{\begin{equation}}
\newcommand{\ee}{\end{equation}}
\newcommand{\bea}{\begin{eqnarray}}
\newcommand{\eea}{\end{eqnarray}}

\begin{document}
\begin{center}
{\Large Condition for confinement in non-Abelian\\ gauge theories}\\
\vspace{1.5em}
{\bf Masud Chaichian\footnote{\email{masud.chaichian@helsinki.fi}} and Marco Frasca\footnote{\email{marcofrasca@mclink.it}}}\\
\vspace{1em}
\textit{$^1$Department of Physics, University of Helsinki\\
 P.O. Box 64, 00014 Helsinki, Finland\\
$^2$Via Erasmo Gattamelata, 3 \\
							00176 Rome, Italy}
\end{center}

\vspace{1em}
\begin{abstract}
We show that a criterion for confinement, based on the BRST invariance, holds in four dimensions, by solving a non-Abelian gauge theory with a set of exact solutions. 
The confinement condition we consider was obtained by Kugo and Ojima some decades ago. 
The current understanding of gauge theories permits us to apply the techniques straightforwardly for checking the validity of this criterion. In this way, we are able to show that the non-Abelian gauge theory is confining and that confinement is rooted in the BRST invariance and asymptotic freedom.
\end{abstract}

\section{Introduction}

The question of why quarks are never seen as single particles is central to a deeper understanding of the Standard Model, especially to the QCD sector, which describes the strong force (\cite{Kogut:2004su} and refs. therein). In the course of years, several mechanisms have been proposed but nobody has been able to derive this property directly from the theory. Rather, some criteria have been devised that can grant confinement in the four dimensional theory. For example, Kugo and Ojima proposed a condition from BRST invariance based on charge annihilation \cite{Kugo:1977zq,Kugo:1979gm}. On a similar ground, Nishijima and collaborators \cite{Nishijima:1993fq,Nishijima:1995ie,Chaichian:2000sf,Chaichian:2005vt,Nishijima:2007ry} obtained a constraint on the amplitudes of unphysical states signaling confinement. These authors showed that colour confinement arises as a consequence of BRST invariance and asymptotic freedom. Indeed, these approaches are deeply linked. In supersymmetric models, confinement is proven in certain conditions as a condensation of monopoles, similar to Type II superconductors \cite{Seiberg:1994aj,Seiberg:1994rs}. For a comparison of different confinement theories and their overlapping regions, see \cite{Chaichian:1999is}. On the other hand, the study of the propagators in the Landau gauge, initiated by Gribov \cite{Gribov:1977wm} and further extended by Zwanziger \cite{Zwanziger:1989mf}, seemed to point to a confining theory with the gluon propagator running to zero as momenta go to zero and an enhanced ghost propagator running to infinity faster than the free case in the same limit of momenta. 

Studies of the gluon and ghost propagators on the lattice, mostly in the Landau gauge, \cite{Bogolubsky:2007ud,Cucchieri:2007md,Oliveira:2007px} and the spectrum \cite{Lucini:2004my,Chen:2005mg} proved that a mass gap appears in a non-Abelian gauge theory without fermions, in evident contrast with the scenario devised by Gribov and Zwanziger. Theoretical support for these results was presented in \cite{Cornwall:1981zr,Cornwall:2010bk,Dudal:2008sp,Frasca:2007uz,Frasca:2009yp,Frasca:2015yva} providing a closed form formula for the gluon propagator. A closed analytical formula for the gluon propagator is pivotal to obtain the low-energy behavior of QCD in a manageable effective theory to prove confinement. Other results are also essential for this aim, as the behavior of the running coupling in the infrared limit \cite{Nesterenko:1999np,Nesterenko:2001st,Nesterenko:2003xb,Baldicchi:2007ic,Baldicchi:2007zn,Bogolubsky:2009dc,Duarte:2016iko} (see also the review \cite{Deur:2016tte}), beside the gluon and ghost propagators. For the latter, the instanton liquid plays an essential role \cite{Schafer:1996wv,Boucaud:2002fx}. Confinement, in its simplest form, can be seen as the combined effect of a potential obtained from the Wilson loop of a Yang--Mills theory without fermions and the running coupling yielding a linear increasing potential, in agreement with lattice data \cite{Deur:2016bwq}. In 2+1 dimensions 
there is no running coupling and the potential increases only logarithmically, granting confinement anyway \cite{Frasca:2016sky}. This latter work shows an exceedingly good agreement with lattice results for the spectrum, giving strong support to the value of this way to solve gauge theories on a lattice.

In this paper, we will apply the condition derived in \cite{Nishijima:1993fq,Nishijima:1995ie,Chaichian:2000sf,Chaichian:2005vt,Nishijima:2007ry}, reducing it to the case of the Kugo--Ojima criterion \cite{Kugo:1979gm}, when, for a non-Abelian gauge theory without fermions, we start with known exact solutions to solve it \cite{Frasca:2015yva}. In this way, the propagators of the theory are known and we can evaluate the behavior of the poles. We will show that this approach permits an explicit computation of the $u$ function of Kugo and Ojima \cite{Kugo:1979gm}.

We point out that our first aim is to consider QCD without quarks, namely  to prove
that a non-Abelian gauge theory with no fermions is confining in four dimensions.
In principle, it provides a rigorous proof that the theory
is confining, besides having a mass gap coming from the derived correlation functions.
At this stage, one can state that confinement is due to the BRST invariance and the asymptotic freedom of the theory, as well the existence of a mass gap.

The paper is structured as follows: in Sec.~\ref{sec2} we introduce the condition for confinement that is obtained from BRST invariance. In Sec.~\ref{sec2a} we present the correlation functions of a non-Abelian gauge theory without fermions, quantized by using a set of exact solutions. In Sec.~\ref{sec2b} we show that the confinement condition is satisfied in this case. In Sec.~\ref{sec2c} we present the exact $\beta$ function. Finally, in Sec.~\ref{sec3} the conclusions are given.

\section{BRST invariance and confinement}
\label{sec2}

In this section we present the approach to confinement proposed in \cite{Nishijima:1993fq,Nishijima:1995ie,Chaichian:2000sf,Chaichian:2005vt,Nishijima:2007ry} and show how this reduces to the Kugo--Ojima criterion \cite{Kugo:1979gm}. We emphasize that our proof is for the theory without fermions.

The Lagrangian of QCD is given by
\begin{equation}
\label{lagrangian}{\cal L}={\cal L}_{{inv}}+{\cal L}_{{gf}}+{\cal
L}_{{FP}},
\end{equation}
where 
${\cal L}_{{inv}}$ denotes the classical gauge-invariant part, ${\cal L}_{{gf}}$ the gauge-fixing terms and ${\cal L}_{{FP}}$ the Faddeev--Popov (FP) ghost term characteristic of non-Abelian gauge theories:
\begin{eqnarray}
\label{eq:L2} 
{\cal L}_{{inv}}&=&-\frac{1}{4}F_{\mu\nu}\cdot
F^{\mu\nu}+\bar\psi(\gamma_\mu D^\mu-m)\psi\,,\cr
{\cal L}_{{gf}}&=&\partial_\mu B\cdot A^\mu+\frac{1}{2}\alpha B\cdot
B\,,\cr
{\cal L}_{{FP}}&=&i\partial_\mu \bar c\cdot D^\mu
c\,,\label{lagr_terms} 
\end{eqnarray}
in the usual notation, with the convention $(1,-1,-1,-1)$ for the metric signature. We denote by $\alpha$ the gauge parameter and $D_\mu$ represents the covariant derivative whose explicit forms are given by
\begin{eqnarray} 
D_\mu\ \psi&=&(\partial_\mu-igT\cdot A_\mu)\psi\,,\cr
 D_\mu\ c&=&\partial_\mu c+gA_\mu\times c\,.\label{cov_deriv}
\end{eqnarray}

In general, the BRST transformations of a generic field $\phi$ are given in terms of the BRST charges $Q_B$ and $\bar Q_B$ by~\cite{Kugo:1977zq}
\begin{eqnarray} 
\delta\,\phi=i[Q_B,\phi]_\mp,\ \ \ \bar\delta\,\phi=i[\bar
Q_B,\phi]_\mp\,,\\\label{brs_phi}
Q_B^2={\bar Q}_B^2=Q_B\bar Q_B+\bar Q_BQ_B=0\,.\label{brs_charge}
\end{eqnarray}
We choose the $-(+)$ sign in (\ref{brs_phi}) when $\phi$ is even (odd) in the ghost fields $c$ and $\bar c$, which are anticommuting scalar fields.

The BRST transformations of the gauge field $A_\mu$ and the quark field $\psi$ are defined by replacing the infinitesimal gauge function by  the FP ghost field $c$ or $\bar c$, in their respective
infinitesimal gauge transformations:
\begin{eqnarray} 
\delta A_\mu&=&D_\mu c\,,\ \ \ \delta\psi=ig(c\cdot
T)\psi\,,\cr
\bar\delta A_\mu&=&D_\mu \bar c\,,\ \ \ \bar\delta\psi=ig(\bar
c\cdot T)\psi\,.\label{brs_transf} 
\end{eqnarray}
Requiring to have for the auxiliary fields $B$, $c$ and $\bar c$ 
\begin{equation} 
\delta{\cal L}=\bar\delta{\cal L}=0\,, 
\end{equation}
we find
\begin{eqnarray} 
\delta\,B=0\,,\ \ \ \delta\,\bar c=i B\,,\ \ \
\delta\,c=-\frac{1}{2}g \,(c\times c)\,,\cr
\bar\delta\,\bar B=0\,,\ \ \ \bar\delta\,c=i \bar B\,,\ \ \
\bar\delta\,\bar c=-\frac{1}{2}g\,(\bar c\times \bar c)\,, 
\end{eqnarray}
where $\bar B$ is defined by
\begin{equation} 
B+\bar B-ig(c\times \bar c)=0\,.
\end{equation}

On the other hand, the conserved current, from Noether theorem (up to a total divergence) is defined as
\begin{equation}
j_{\mu}=\sum_{\{\Phi\}}\frac{\partial \ L}{\partial (\partial_\mu \Phi)}\delta\Phi
=B^a(D_\mu c)^a -\partial_\mu B^a c^a+i\frac12{\rm g}f^{abc}\partial_\mu \bar c^a c^bc^c,
\end{equation}
with $\{\Phi\}$ the set of all fields present in the Lagrangian, from which we get the corresponding charge $Q_B$:
\begin{equation}
Q_B=\int d^3x \left(B^a(D_0 c)^a -\dot B^a c^a+i\frac12{\rm g}f^{abc}\dot{\bar c}^a c^bc^c\right).
\end{equation}

So, the Lagrangian with the gauge-fixing term is then
\begin{equation}
{\cal L}_{{gf}}+{\cal L}_{{FP}}=\delta(-i\partial_\mu\bar c\cdot
A_\mu-\frac{i}{2}\alpha\,\bar c \cdot B)
\end{equation}
and evidently we have
\begin{equation} 
\delta{\cal L}_{inv}=0\,. 
\end{equation}
Namely, ${\cal L}_{inv}$ is closed and ${\cal L}_{{gf}}+{\cal L}_{{FP}}$ is exact, and
\begin{equation} 
\delta{\cal L}=0\,. 
\end{equation}

This Lagrangian yields the equations of motion
\begin{equation}
   D^{\mu ab}F^b_{\mu\nu}+j^b_\nu=i\delta\bar\delta A^b_\nu,
\end{equation}
where the contribution on the right-hand side comes from the auxiliary fields in the Lagrangian. At the tree level, these represent massless particles. Besides, the $B$ field does not propagate. This means that the current due to these fields should not yield contributions to the physical spectrum of the theory. Also, since $\partial^\nu(i\delta\bar\delta A_\nu)=0$, this current is conserved. In order to evaluate it, we need to study the behavior of the amplitude
\begin{equation}
\label{eq:corr}
\begin{array}{lll}
\langle i \delta \bar{\delta} A_{\mu} ^{a} (x), A_{\nu} ^{b} (y)
\rangle.
\end{array}
\end{equation} \noindent Then, according to current conservation, the most general form of its Fourier transform can be expressed as

\begin{equation}
\label{eq:KO}
\begin{array}{lll}
-\delta^{ab}(\delta_{\mu\nu} - \frac{p_{\mu} p_{\nu}}{p^2+i\epsilon}) \int
dm^2 \frac{\sigma (m^2)}{p^2-m^2+i\epsilon} + C
\delta^{ab}\frac{p_{\mu}p_{\nu}}{p^2+i\epsilon}.
\end{array}
\end{equation} \noindent As we see, we cannot exclude massless excitations from the spectrum at this stage. This will imply no confinement, as we would get massless gluons. But if the theory is confining, massless states cannot be physical states. Then,

\begin{equation}
\begin{array}{lll}
\partial _{\mu} \langle i \delta \bar{\delta} A_{\mu}^{a} (x),
A_{\nu}^{b} (y) \rangle = i \delta_{ab} C \partial _{\nu} \delta
^4 (x-y),
\end{array}
\end{equation} 
\noindent that can be cast into the form of an equal-time commutator:
\begin{equation}
\label{eq:comm}
\begin{array}{lll}
\delta (x_0-y_0) \langle 0 \vert \left[i\delta \bar{\delta}
A_0^{a} (x), A_j^{b} (y)\right] \vert 0 \rangle& = &i \delta_{ab}
C \partial _j \delta ^4 (x-y),\\&& (j=1,2,3).
\end{array}
\end{equation}

Based on the preceding considerations, we have seen that the confinement condition is realized with 
no massless excitations appearing in the physical spectrum and the current arising from the auxiliary fields has no effect on the amplitudes of the processes. 

We can link this conclusion with the Kugo--Ojima criterion, which is also a no-massless pole criterion. This can be seen in the following way. 
Using the Kugo--Ojima formalism, one has,
\begin{equation}
   \delta \bar{\delta} A_{\mu}^{a}=-\{Q_B,\{\bar{Q}_B,A_{\mu} ^{a}\}\}.
\end{equation}
Using the fact that $\langle 0|Q_B=Q_B|0\rangle=\bar{Q}_B|0\rangle=\langle 0|\bar{Q}_B=0$, it is clear that
\begin{equation}
   \langle i \delta \bar{\delta} A_{\mu}^{a} (x),A_{\nu}^{b} (y) \rangle=\langle i  \bar{\delta} A_{\mu}^{a} (x),\delta A_{\nu}^{b} (y) \rangle=
	i\langle D_\mu\bar{c}^{a} (x),D_\nu c^{b} (y) \rangle.
\end{equation}
For this correlator, Kugo and Ojima showed  \cite{Kugo:1979gm} that
\begin{equation}
\label{eq:KOc}
\int d^dxe^{ipx}\langle D_\mu\bar{c}^{a} (x),D_\nu c^{b} (y) \rangle=\delta^{ab}
\left(\delta_{\mu\nu} - \frac{p_{\mu} p_{\nu}}{p^2-i\epsilon}\right) u(p^2)-\delta^{ab}\frac{p_{\mu} p_{\nu}}{p^2-i\epsilon},
\end{equation}
and the no-pole condition
yields here
\begin{equation} 
1+u(p^2=0)=0,
\end{equation}
which is the Kugo--Ojima condition for confinement. 

Thus, our aim will be to derive the $u(p^2)$ function and evaluate it for $p^2=0$.

At this stage we note that a possible mapping exists between the Nishijima condition and the Kugo--Ojima condition when the infrared limit $p^2\rightarrow 0$ is taken. From eq. (\ref{eq:KO}) we get
\begin{eqnarray}
-\delta^{ab}\left(\delta_{\mu\nu} - \frac{p_{\mu} p_{\nu}}{p^2+i\epsilon}\right) \int
dm^2 \frac{\sigma (m^2)}{p^2-m^2+i\epsilon} + C
\delta^{ab}\frac{p_{\mu}p_{\nu}}{p^2+i\epsilon}\cr
\stackrel{p^2\rightarrow 0}{\rightarrow}
\delta^{ab}\left(\delta_{\mu\nu} - \frac{p_{\mu} p_{\nu}}{p^2+i\epsilon}\right) \int
dm^2 \frac{\sigma (m^2)}{m^2} + C
\delta^{ab}\frac{p_{\mu}p_{\nu}}{p^2+i\epsilon}.
\end{eqnarray}
On the other hand, the no-massless pole condition must be taken into account as
\begin{equation}
   C-\int dm^2 \frac{\sigma (m^2)}{m^2}=0.
\end{equation}
This is analogous to the Kugo--Ojima condition $1+u(p^2=0)=0$ in the infrared limit.  

\section{Correlation functions in a non-Abelian gauge theory\label{sec2a}}

The correlation functions for a pure non-Abelian gauge theory, without matter fields, have been computed in \cite{Frasca:2015yva}, where the Dyson--Schwinger equations were solved with the approach devised in \cite{Bender:1999ek}. In these computations, the Dyson--Schwinger equations are solved with no truncation involved but computations are performed to obtain at least the two-point function exactly. For the sake of completeness, we give a summary of them in the appendix. Below, we present the solutions.

We note that $G_{1\mu}^a(x)$ can be written as in (\ref{eq:exact})
\begin{equation}
   G_{1\mu}^a(x)=\eta^a_\mu\phi(x),
\end{equation}
where $\phi(x)=\mu\left(\frac{2}{Ng^2}\right)^\frac{1}{4}\cdot{\rm sn}(px,-1)$, with $\eta^a_\mu$ constants and $p^2=\mu^2\sqrt{Ng^2/2}$. Thus,
the given set of Dyson--Schwinger equations can be solved exactly. For the two-point function in the Landau gauge we can write
\begin{equation}
    G^{ab}_{\mu\nu}(x)=\delta_{ab}\left(g_{\mu\nu}-\frac{p_\mu p_\nu}{p^2}\right)\Delta(x-y),
\end{equation}
provided that
\begin{eqnarray}
\label{eq:prop}
    &&\partial^2\Delta(x-y)+3Ng^2\phi^2(x)\Delta(x-y)=\delta^4(x-y), \nonumber \\
		&&P_1^a(x)=0, \nonumber \\
	 &&\partial^2 P^{am}_2(x-y)=\delta_{am}\delta^4(x-y), \nonumber \\
		&&K^{am}_{2\kappa}(x-y)=0 
\end{eqnarray}
and $G_{2\nu\rho}^{ac}(0)=0$, $G_{3\mu\nu\kappa}^{bcm}(0,x-y)=0$, $G_{4\mu\nu\kappa}^{\mu bdem}(0,0,x-y)$, $K^{bcm}_{3\kappa}(0,x-y)=0$, a behavior of the 3- and 4-point functions in agreement with lattice results \cite{Cucchieri:2008qm,Duarte:2016ieu}. This shows that the set of Schwinger--Dyson equations for Yang--Mills theory can be exactly solved, at least to the level of two-point functions. 

The propagator is given by \cite{Frasca:2015yva}
\begin{equation}
   \Delta(p)=\frac{\pi^3}{4K^3(-1)}
	\sum_{n=0}^\infty\frac{e^{-(n+\frac{1}{2})\pi}}{1+e^{-(2n+1)\pi}}(2n+1)^2\frac{1}{p^2-m_n^2+i\epsilon},
\end{equation}
with $K(-1)$ being an elliptic integral that yields the numerical constant $1.3110287771460598\ldots$ and given the mass spectrum
\begin{equation}
\label{eq:spec}
   m_n=(2n+1)\frac{\pi}{2K(-1)}\left(\frac{Ng^2}{2}\right)^\frac{1}{4}\sigma_0^\frac{1}{2},
\end{equation}
that is indeed the spectrum of the theory. Here $\sigma_0$ is an integration constant having the dimension of mass. It is easy to see how this propagator recovers asymptotic freedom \cite{Nishijima:1993fq,Nishijima:1995ie,Chaichian:2000sf,Chaichian:2005vt,Nishijima:2007ry}. In the high-energy limit, we make the momenta run to infinity. This yields
\begin{equation}
   \Delta(p)\stackrel{p\rightarrow\infty}{=} \frac{\pi^3}{4K^3(-1)}
	\sum_{n=0}^\infty\frac{e^{-(n+\frac{1}{2})\pi}}{1+e^{-(2n+1)\pi}}(2n+1)^2 p^{-2}=p^{-2},
\end{equation}
as the sum adds to 1. We just note that this propagator is a leading order approximation when one can neglect the corrections due to mass renormalization to the spectrum of the theory.

The theory has no massless excitation and thus, already at this stage, we can conclude that the approach devised in \cite{Kugo:1977zq,Kugo:1979gm,Nishijima:1993fq,Nishijima:1995ie,Chaichian:2000sf,Chaichian:2005vt,Nishijima:2007ry} appears sound. We will complete the proof in the next section.

\section{Confinement condition\label{sec2b}}

Now, we are in a position to evaluate the confinement condition by computing the $u(p^2)$ function and evaluating it at 0. For the sake of simplicity we limit our analysis to $SU(N)$ and the numerical analysis to $SU(3)$. This extends the analysis, performed on the lattice, presented in \cite{Sternbeck:2006rd, Aguilar:2009pp}. 
We note that, from eq. (\ref{eq:KOc}),
\begin{equation}
\int d^4xe^{ipx}\langle D_\mu\bar{c}^{a} (x),D_\nu c^{b} (0) \rangle=
\int d^4xe^{ipx}\langle\left(\partial_\mu-igT^cA^c_\mu(x)\right)\bar{c}^{a} (x),
\left(\partial_\nu-igT^dA^d_\nu(0)\right)c^{b} (0)\rangle.
\end{equation}
This yields
\begin{equation}
\int d^4xe^{ipx}\langle D_\mu\bar{c}^{a} (x),D_\nu c^{b} (0) \rangle=-\delta^{ab}\frac{p_\mu p_\nu}{p^2}-g^2
\int d^4xe^{ipx}\langle T^cA^c_\mu(x)\bar{c}^a(x),T^dA^d_\nu(0)c^b(0)\rangle,
\end{equation}
where it has been taken into account that $\langle A^a_\mu(x)\rangle=0$ and we used the free ghost propagator. Now, as shown in the preceding section, the ghost field decouples from the gluon field and the above equation can be written as follows:
\begin{eqnarray}
\int d^4xe^{ipx}\langle D_\mu\bar{c}^{a} (x),D_\nu c^{b} (y) \rangle&=&-\delta^{ab}\frac{p_\mu p_\nu}{k^2}\\
&-&\frac{(N^2-1)^2}{2N}g^2\delta^{ab}
\left(\delta_{\mu\nu} - \frac{p_{\mu} p_{\nu}}{p^{2}}\right)
\int\frac{d^4p'}{(2\pi)^4}\frac{1}{|p-p'|^2}\Delta(p'),\nonumber
\end{eqnarray}
where we identify
\begin{equation}
  u(p^2)=-\frac{(N^2-1)^2}{2N}g^2\int\frac{d^4p'}{(2\pi)^4}\frac{1}{|p-p'|^2}\Delta(p').
\end{equation}
Then, we have to evaluate the integral
\begin{eqnarray}
  u(0)&=&-\frac{(N^2-1)^2}{2N}g^2\int\frac{d^4p}{(2\pi)^4}\frac{1}{p^2}\sum_{n=0}^\infty B_n\frac{1}{p^2+m_n^2}\cr
&=&
	-\frac{(N^2-1)^2}{2N}g^2\sum_{n=0}^\infty\frac{B_n}{m_n^2}
	\int\frac{d^4p}{(2\pi)^4}\left(\frac{1}{p^2}-\frac{1}{p^2+m_n^2}\right),
\end{eqnarray}
with $B_n=\frac{\pi^3}{4K^3(-1)}\frac{e^{-(n+\frac{1}{2})\pi}}{1+e^{-(2n+1)\pi}}(2n+1)^2$. This integral is divergent and needs to be renormalized. We can evaluate it by dimensional regularization. We use
\begin{equation}
  I_d=-\int\frac{d^dp}{(2\pi)^d}\left(\frac{1}{p^2}-\frac{1}{p^2+m_n^2}\right)=\frac{(m_n^2)^{d/2-1}}{(4\pi)^\frac{d}{2}}\Gamma(1-d/2),
\end{equation}
then set $\epsilon=4-d$ and expand. This yields
\begin{equation}
  I_\epsilon =\frac{m_n^2}{(4\pi)^2}\left(\frac{4\pi\mu^2}{m_n^2}\right)^\frac{\epsilon}{2}\Gamma\left(\frac{\epsilon}{2}-1\right)=
	\frac{m_n^2}{(4\pi)^2}\left[-\frac{2}{\epsilon}-1+\gamma+\ln\left(\frac{m_n^2}{4\pi\mu^2}\right)+O(\epsilon)\right],
\end{equation}
where we have reintroduced the scale factor $\mu$ arising by going to $d$ dimensions and $\gamma$ is the Euler--Mascheroni constant. From this we can extract the finite part, that is
\begin{equation}
  I'_4=\frac{m_n^2}{(4\pi)^2}\left[-1+\gamma+\ln\left(\frac{m_n^2}{4\pi\mu^2}\right)\right],
\end{equation}
which is explicitly dependent on the cut-off $\mu$. Then,
\begin{equation}
\label{eq:u0}
  u(0)=\frac{(N^2-1)^2}{2N}\frac{\alpha_s}{4\pi}\left[-1+\gamma+\sum_{n=0}^\infty B_n\ln\left(\frac{m_n^2}{4\pi\mu^2}\right)\right],
\end{equation}
where use has been made of the identity $\sum_{n=0}^\infty B_n=1$ and $\alpha_s=g^2/4\pi$. 

One can see that, if for the Kugo--Ojima function holds $u(0)=-1$ granting confinement, we obtain a running coupling $\alpha_s(\mu^2)$ given by the following equation
\begin{equation}
\label{eq:rc}
    \frac{(N^2-1)^2}{2N}\frac{\alpha_s(\mu^2)}{4\pi}\left[-1+\gamma+\sum_{n=0}^\infty B_n\ln\left(\frac{m_n^2}{4\pi\mu^2}\right)\right]=-1.
\end{equation}
This equation, consistently with our approach, is exact. Indeed, in the high-energy limit, we get the asymptotic freedom limit for $SU(3)$ as
\begin{equation}
\label{eq:alpha1}
   \alpha_s(\mu^2)=\frac{3\pi}{8\ln\left(\frac{\mu^2}{\sigma}\right)},
\end{equation}
where use has been made of eq. (\ref{eq:spec}) for the spectrum of the theory and we have introduced the string tension $\sigma=(0.44\ \mbox{MeV})^2$ obtained from experimental data that we keep here fixed. The square root of the string tension represents the gap into the spectrum of the theory and, when one accounts for quarks, characterizes the glueball spectrum. This result should compare with the asymptotic freedom limit given by \cite{PDG}
\begin{equation}
\label{eq:alpha2}
   \alpha_s(\mu^2)=\frac{12\pi}{(33-2n_f)\ln\left(\frac{\mu^2}{\sigma}\right)},
\end{equation}
with $n_f$ being the number of flavours of quarks that here we take to be 0 and we assume $\sigma$ also here for the integration constant coming from the equation of the renormalization group. This is just for reasons of numerical comparison but we note that it is physically meaningful anyway. In this way, one gets the ratio between eqs.(\ref{eq:alpha2}) and (\ref{eq:alpha1}) equal to $96/99\approx 0.97$, very near 1, but we should remember that the former is a perturbative result in an asymptotic series.

We can also compare with the experimental value of $\alpha_s$ at $M_Z$, the mass of the Z vector boson. From \cite{PDG} one has the world average value $\alpha_s(M_Z^2)=0.1181\pm 0.0011$ while our result is $\alpha_s(M_Z^2)=0.110\pm 0.005$, having estimated an error of 22 MeV on $\sqrt{\sigma}$. The agreement is within an error of about 7\%. We have not accounted contribution of quarks in this computation. We just note that the analogous limit from perturbative QCD has a higher error. Also, the perturbative result is very near to this value being about $0.107\pm 0.005$.

\section{$\beta$ function \label{sec2c}}

So far, we have evaluated the running coupling, given by eq. (\ref{eq:rc}), fixing the gap in the spectrum of the theory, given by eq. (\ref{eq:spec}). This requires solving the eq. (\ref{eq:rc}) by iteration. Notwithstanding, this yields excellent results for asymptotic freedom; we need to see if this agreement will extend for all the energy range. This can be done by deriving the $\beta$ function from eq. (\ref{eq:rc}) without any approximation. We do it by noting that the spectrum depends on $\alpha_s$ and, normally, we set for the string tension (the gap in the spectrum)
\begin{equation}
   \sigma = \sigma_0\sqrt{2\pi N\alpha_s}.
\end{equation}
The idea is to use $\sigma_0$ as an energy scale for the ultraviolet cut-off $\mu$ we introduced in the preceding section after renormalization of the $u$ function. Given this, we can derive the $\beta$ function from eq. (\ref{eq:rc}) in a straightforward manner. This gives the renormalization group equation
\begin{equation}
   \frac{d\alpha_s}{dl}=-\beta_0\frac{\alpha_s^2}{1-\frac{1}{2}\beta_0\alpha_s},
\end{equation}
with $\beta_0=(N^2-1)^2/8\pi N$. We have set $l=\ln(\mu^2/\sigma_0)$ as an independent variable. This result should compare with the exact $\beta$ function obtained for SUSY Yang--Mills theory \cite{Novikov:1983uc,Shifman:1986zi}
\begin{equation}
\label{eq:SUSY}
   \frac{d\alpha_s}{dl}=-\frac{3N}{4\pi}\frac{\alpha_s^2}{1-\frac{1}{2\pi}N\alpha_s},
\end{equation}
and the Ryttov and Sannino hypothesis for Yang--Mills theory \cite{Ryttov:2007cx}
\begin{equation}
   \frac{d\alpha_s}{dl}=-\frac{3N}{12\pi}\frac{\alpha_s^2}{1-\frac{34}{44\pi}N\alpha_s}.
\end{equation}
It should be pointed out that the Ryttov--Sannino hypothesis, also being inspired by the SUSY result of eq. (\ref{eq:SUSY}), is founded on the perturbative results of asymptotic freedom as given in \cite{PDG}.

It is interesting to note that, in the formal limit $\alpha_s\rightarrow\infty$, SUSY Yang--Mills theory gives for the $\beta$ function $3\alpha_s/2$ while our equation yields $2\alpha_s$ in the same limit. However, Ryttov and Sannino would get about $0.3\alpha_s$ in the same limit.

\section{Conclusions}
\label{sec3}

Using the approaches developed in \cite{Kugo:1977zq,Kugo:1979gm} and \cite{Nishijima:1993fq,Nishijima:1995ie,Chaichian:2000sf,Chaichian:2005vt,Nishijima:2007ry}, we were able to give a rigorous proof of confinement for non-Abelian gauge theories in four dimensions as a consequence of the BRST invariance and the asymptotic freedom. Our results are based on the exact solutions obtained in \cite{Frasca:2015yva} for the correlation functions. These are obtained by solving the set of Schwinger--Dyson equations exactly, without truncation, to obtain the exact two-point function. As a by-product, we get an exact equation for the running coupling of the theory. 

We hope to extend this proof to the case of QCD with fermions in a future communication.

\section*{Acknowledgements}

We are deeply indebted to Taichiro Kugo for several enlightening discussions and comments on the manuscript, which have improved the results and the conclusions of the work to a significant degree. Our thanks also go to Carl Bender for enlightening discussions and to David Dudal and Silvio Sorella for pointing to us several originally weak points, where the exact solutions were confronted with the lattice simulations and as well for useful suggestions. Last but not least, we are grateful to Marco Ruggieri for useful discussions about the various questions discussed in the work.

\appendix
\section*{Appendix: Dyson--Schwinger equations \label{app1}}

The correlation functions are obtained when a given exact solution is known for the one-point function i.e., one has to solve exactly the equations
\begin{equation}
\partial^\mu\partial_\mu A^a_\nu-\frac{1}{2\alpha}\partial_\nu(\partial^\mu A^a_\mu)+gf^{abc}A^{b\mu}(\partial_\mu A^c_\nu-\partial_\nu A^c_\mu)+gf^{abc}\partial^\mu(A^b_\mu A^c_\nu)+g^2f^{abc}f^{cde}A^{b\mu}A^d_\mu A^e_\nu = 0.
\end{equation} 
In the Landau gauge ($\alpha\rightarrow 0$), these are exactly given in the form
\begin{equation}
\label{eq:exact}
   A^a_\nu(x)=\eta^a_\nu\left(\frac{2}{Ng^2}\right)^\frac{1}{4}\mu\cdot{\rm sn}(px,-1),
\end{equation}
with ${\rm sn}(px,-1)$ the Jacobi snoidal elliptic function and $\eta_\mu^a$ being a set of constants to be determined depending on the problem at hand (e.g., for $SU(2)$ one can take $\eta_1^1=\eta_2^2=\eta_3^3=1$, all other components being zero) and $\mu$ an integration constant with the dimension of an energy. This holds provided the following dispersion relation holds
\begin{equation}
\label{eq:disp}
    p^2=\sqrt{\frac{Ng^2}{2}}\mu^2.
\end{equation}
Solutions given in eq.(\ref{eq:exact}) appear as massive solution, due to the dispersion relation (\ref{eq:disp}), even if we started from a massless theory. 

Then, if we use these solutions as one-point function of the set of Schwinger--Dyson equations for a non-Abelian gauge theory without fermions, given by \cite{Frasca:2015yva}, we are able to compute the two-point functions exactly, without any approximation or truncation. We use the approach devised in \cite{Bender:1999ek}. Indeed, to get the Schwinger--Dyson equations one has to start from the quantum equations of motion that have the form
\begin{eqnarray}
   &&\partial^\mu\partial_\mu A^a_\nu+gf^{abc}A^{b\mu}(\partial_\mu A^c_\nu-\partial_\nu A^c_\mu)+gf^{abc}\partial^\mu(A^b_\mu A^c_\nu)+g^2f^{abc}f^{cde}A^{b\mu}A^d_\mu A^e_\nu \nonumber \\
	&&= gf^{abc}\partial_\nu(\bar c^b c^c) + j_\nu^a, \nonumber \\
	 &&\partial^\mu\partial_\mu c^a+gf^{abc}\partial^\mu(A_\mu^bc^c)=\varepsilon^a. 
\end{eqnarray}
We fix the gauge to the Landau gauge, $\alpha\rightarrow 0$, and $c,\ \bar c$ are the ghost fields. Averaging on the vacuum state and dividing by the partition function $Z_{YM}[j,\bar\varepsilon,\varepsilon]$, one has
\begin{eqnarray}
    &&\partial^2G_{1\nu}^{(j)a}(x)+
		gf^{abc}(\langle A^{b\mu}\partial_\mu A^c_\nu\rangle-\langle A^{b\mu}\partial_\nu A^c_\mu\rangle)Z^{-1}_{YM}[j,\bar\varepsilon,\epsilon]
		+gf^{abc}\partial^\mu\langle A^b_\mu A^c_\nu\rangle Z^{-1}_{YM}[j,\bar\varepsilon,\varepsilon]
		\nonumber \\
		&&+g^2f^{abc}f^{cde}\langle A^{b\mu}A^d_\mu A^e_\nu\rangle Z^{-1}_{YM}[j,\bar\varepsilon,\varepsilon] 
		=gf^{abc}\langle\partial_\nu(\bar c^b c^c)\rangle Z^{-1}_{YM}[j,\bar\varepsilon,\varepsilon] + j_\nu^a, \nonumber \\
	 &&\partial^2 P^{(\varepsilon)a}_1(x)
	 +gf^{abc}\partial^\mu\langle A_\mu^bc^c\rangle Z^{-1}_{YM}[j,\bar\varepsilon,\varepsilon]=\varepsilon^a. 
\end{eqnarray}
The one-point function is given by
\begin{eqnarray}
    &&G_{1\nu}^{(j)a}(x)Z_{YM}[j,\bar\varepsilon,\epsilon]=\langle A^a_\nu(x)\rangle, \nonumber \\
		&&P^{(\varepsilon)a}_1(x)Z_{YM}[j,\bar\varepsilon,\epsilon]=\langle c^a(x)\rangle. 
\end{eqnarray}
Deriving once with respect to currents, at the same point because of the averages on the vacuum (see \cite{Bender:1999ek}), one has
\begin{eqnarray}
   &&G_{2\nu\kappa}^{(j)ab}(x,x)Z_{YM}[j,\bar\varepsilon,\epsilon]+G_{1\nu}^{(j)a}(x)G_{1\kappa}^{(j)b}(x)Z_{YM}[j,\bar\varepsilon,\epsilon]=\langle A^a_\nu(x)A^b_\kappa(x)\rangle, \nonumber \\
	 &&P^{(\varepsilon)ab}_2(x,x)Z_{YM}[j,\bar\varepsilon,\epsilon]+\bar P^{(\varepsilon)a}_1(x)P^{(\varepsilon)b}_1(x)Z_{YM}[j,\bar\varepsilon,\epsilon]=
	\langle \bar c^b(x)c^a(x)\rangle, \nonumber \\
	&&\partial_\mu G_{2\nu\kappa}^{(j)ab}(x,x)Z_{YM}[j,\bar\varepsilon,\epsilon]+\partial_\mu G_{1\nu}^{(j)a}(x)G_{1\kappa}^{(j)b}(x)Z_{YM}[j,\bar\varepsilon,\epsilon]=
	\langle\partial_\mu A^a_\nu(x)A^b_\kappa(x)\rangle, \nonumber \\
	&&K^{(\varepsilon,j)ab}_{2\nu}(x,x)Z_{YM}[j,\bar\varepsilon,\epsilon]+P^{(\varepsilon)a}_1(x)G_{1\nu}^{(j)b}(x)Z_{YM}[j,\bar\varepsilon,\epsilon]=\langle c^a(x)A_\nu^b(x)\rangle,
\end{eqnarray}
and twice
\begin{eqnarray}
   &&G_{3\nu\kappa\rho}^{(j)abc}(x,x,x)Z_{YM}[j,\bar\varepsilon,\epsilon]+G_{2\nu\kappa}^{(j)ab}(x,x)G_{1\rho}^{(j)c}(x)Z_{YM}[j,\bar\varepsilon,\epsilon]+ \nonumber \\
	&&G_{2\nu\rho}^{(j)ac}(x,x)G_{1\kappa}^{(j)b}(x)Z_{YM}[j,\bar\varepsilon,\epsilon]
	+G_{1\nu}^{(j)a}(x)G_{2\kappa\rho}^{(j)bc}(x,x)Z_{YM}[j,\bar\varepsilon,\epsilon]+ \nonumber \\
	&&G_{1\nu}^{(j)a}(x)G_{1\kappa}^{(j)b}(x)G_{1\rho}^{(j)c}(x)Z_{YM}[j,\bar\varepsilon,\epsilon]=\langle A^a_\nu(x)A^b_\kappa(x)A^c_\rho(x)\rangle.
\end{eqnarray}
These give us the first set of Schwinger--Dyson equations as
\begin{eqnarray}
\label{eq:ds_1}
    &&\partial^2G_{1\nu}^{(j)a}(x)+gf^{abc}(
		\partial^\mu G_{2\mu\nu}^{(j)bc}(x,x)+\partial^\mu G_{1\mu}^{(j)b}(x)G_{1\nu}^{(j)c}(x)-
		\partial_\nu G_{2\mu}^{(j)\mu bc}(x,x)-\partial_\nu G_{1\mu}^{(j)b}(x)G_{1}^{(j)\mu c}(x)) \nonumber \\
		&&+gf^{abc}\partial^\mu G_{2\mu\nu}^{(j)bc}(x,x)+gf^{abc}\partial^\mu(G_{1\mu}^{(j)b}(x)G_{1\nu}^{(j)c}(x))		
		\nonumber \\
		&&+g^2f^{abc}f^{cde}(G_{3\mu\nu}^{(j)\mu bde}(x,x,x)
		+G_{2\mu\nu}^{(j)bd}(x,x)G_{1}^{(j)\mu e}(x) \nonumber \\
	&&+G_{2\nu\rho}^{(j)eb}(x,x)G_{1}^{(j)\rho d}(x)
	+G_{2\mu\nu}^{(j)de}(x,x)G_{1}^{(j)\mu b}(x)+ \nonumber \\
	&&G_{1}^{(j)\mu b}(x)G_{1\mu}^{(j)d}(x)G_{1\nu}^{(j)e}(x))
		=gf^{abc}(\partial_\nu P^{(\varepsilon)bc}_2(x,x)+\partial_\nu (\bar P^{(\varepsilon)b}_1(x)P^{(\varepsilon)c}_1(x))) 
		+ j_\nu^a, \nonumber \\
	 &&\partial^2 P^{(\varepsilon)a}_1(x)+gf^{abc}\partial^\mu
	(K^{(\varepsilon,j)bc}_{2\mu}(x,x)+P^{(\varepsilon)b}_1(x)G_{1\mu}^{(j)c}(x))=\varepsilon^a. 
\end{eqnarray}
By setting the currents to zero and noticing that, by translation invariance, one has $G_2(x,x)=G_2(x-x)=G_2(0)$, $G_3(x,x,x)=G_3(0,0)$ and $K_2(x,x)=K_2(0)$, we get
\begin{eqnarray}
    &&\partial^2G_{1\nu}^{a}(x)+gf^{abc}(
		\partial^\mu G_{2\mu\nu}^{bc}(0)+\partial^\mu G_{1\mu}^{b}(x)G_{1\nu}^{c}(x)-
		\partial_\nu G_{2\mu}^{\nu bc}(0)-\partial_\nu G_{1\mu}^{b}(x)G_{1}^{\mu c}(x)) \nonumber \\
		&&+gf^{abc}\partial^\mu G_{2\mu\nu}^{bc}(0)+gf^{abc}\partial^\mu(G_{1\mu}^{b}(x)G_{1\nu}^{c}(x))		
		\nonumber \\
		&&+g^2f^{abc}f^{cde}(G_{3\mu\nu}^{\mu bde}(0,0)
		+G_{2\mu\nu}^{bd}(0)G_{1}^{\mu e}(x) \nonumber \\
	&&+G_{2\nu\rho}^{eb}(0)G_{1}^{\rho d}(x)
	+G_{2\mu\nu}^{de}(0)G_{1}^{\mu b}(x)+ \nonumber \\
	&&G_{1}^{\mu b}(x)G_{1\mu}^{d}(x)G_{1\nu}^{e}(x))
		=gf^{abc}(\partial_\nu P^{bc}_2(0)+\partial_\nu (\bar P^{b}_1(x)P^{c}_1(x))), \nonumber \\
	 &&\partial^2 P^{a}_1(x)+gf^{abc}\partial^\mu
	(K^{bc}_{2\mu}(0)+P^{b}_1(x)G_{1\mu}^{c}(x))=0. 
\end{eqnarray}
This set of Schwinger--Dyson equations can be solved exactly in the Landau gauge with the aforementioned exact solutions. 
This is so by noting that the contributions coming from $G_{2\mu\nu}^{ab}(0)$, $P_2^{ab}(0)$, $G_{3\mu\nu}^{\mu bde}(0,0)$ and $K^{bc}_{2\mu}(0)$ are zero in this case due to the fact that they give a symmetric group contribution against the antisymmetric structure constants of the group itself. Then, one gets that the ghost one-point function decouples and can be assumed to be a constant and does not contribute to the gluon one-point function.

The Schwinger--Dyson equation for the two-point functions can be obtained by further deriving eq. (\ref{eq:ds_1}). One has
\begin{eqnarray}
\label{eq:ds_2}
    &&\partial^2G_{2\nu\kappa}^{(j)am}(x-y)
		+gf^{abc}(
		\partial^\mu G_{3\mu\nu\kappa}^{(j)bcm}(x,x,y)
		+\partial^\mu G_{2\mu\kappa}^{(j)bm}(x-y)G_{1\nu}^{(j)c}(x)
		+\partial^\mu G_{1\mu}^{(j)b}(x)G_{2\nu\kappa}^{(j)cm}(x-y) \nonumber \\
		&&-\partial_\nu G_{3\mu\kappa}^{(j)\mu bcm}(x,x,y)
		-\partial_\nu G_{2\mu\kappa}^{(j)bm}(x-y)G_{1}^{(j)\mu c}(x)) 
		-\partial_\nu G_{1\mu}^{(j)b}(x)G_{2\kappa}^{(j)\mu cm}(x-y))
		\nonumber \\
		&&+gf^{abc}\partial^\mu G_{3\mu\nu\kappa}^{(j)bcm}(x,x,y)
		+gf^{abc}\partial^\mu(G_{2\mu\kappa}^{(j)bm}(x-y)G_{1\nu}^{(j)c}(x))
				+gf^{abc}\partial^\mu(G_{1\mu}^{(j)b}(x)G_{1\nu\kappa}^{(j)cm}(x-y))
		\nonumber \\
		&&+g^2f^{abc}f^{cde}(G_{4\mu\nu\kappa}^{(j)\mu bdem}(x,x,x,y)
		+G_{3\mu\nu\kappa}^{(j)bdm}(x,x,y)G_{1}^{(j)\mu e}(x) 
		+G_{2\mu\nu}^{(j)bd}(x,x)G_{2\kappa}^{(j)\mu em}(x-y)\nonumber \\
	&&+G_{3\nu\rho\kappa}^{(j)acm}(x,x,y)G_{1}^{(j)\rho b}(x)
	+G_{2\nu\rho}^{(j)eb}(x,x)G_{2\kappa}^{(j)\rho dm}(x-y) \nonumber \\
	&&+G_{2\nu\rho}^{(j)de}(x,x)G_{2\kappa}^{(j)\rho bm}(x-y)
	+G_{1}^{(j)\mu b}(x)G_{3\mu\nu\kappa}^{(j)dem}(x,x,y)+ \nonumber \\
	&&G_{2\kappa}^{(j)\mu bm}(x-y)G_{1\mu}^{(j)d}(x)G_{1\nu}^{(j)e}(x)+
	G_{1}^{(j)\mu b}(x)G_{2\mu\kappa}^{(j)dm}(x-y)G_{1\nu}^{(j)e}(x)+
	G_{1}^{(j)\mu b}(x)G_{1\mu}^{(j)d}(x)G_{2\nu\kappa}^{(j)em}(x-y)) \nonumber \\
	&&	=gf^{abc}(\partial_\nu K^{(j\varepsilon)bcm}_{3\kappa}(x,x,y)
	+\partial_\nu (\bar P^{(\varepsilon)b}_1(x)K^{(j\varepsilon)cm}_{2\kappa}(x,y))) \nonumber \\
	&&+\partial_\nu (\bar K^{(j\varepsilon)bm}_{2\kappa}(x,y)P^{(\varepsilon)c}_1(x))) 
	+ \delta_{am}g_{\nu\kappa}\delta^4(x-y), \nonumber \\
	 &&\partial^2 P^{(\varepsilon)am}_2(x-y)+gf^{abc}\partial^\mu
	(K^{(\varepsilon,j)bcm}_{3\mu}(x,x,y)+P^{(\varepsilon)bm}_2(x-y)G_{1\mu}^{(j)c}(x)+ \nonumber \\
	&&P^{(\varepsilon)b}_1(x)K_{2\mu}^{(j\varepsilon)cm}(x-y))=\delta_{am}\delta^4(x-y), \nonumber \\
	&&\partial^2 K^{(j\varepsilon)am}_{2\kappa}(x-y)+gf^{abc}\partial^\mu
	(L^{(\varepsilon,j)bcm}_{2\mu\kappa}(x,x,y)+ \nonumber \\
	&&K^{(j\varepsilon)bm}_{2\kappa}(x-y)G_{1\mu}^{(j)c}(x)+P^{(\varepsilon)b}_1(x)G_{2\mu\kappa}^{(j)cm}(x-y))=0. 
\end{eqnarray}
By setting currents to zero and using translation invariance, the above mentioned relations yield
\begin{eqnarray}
\label{eq:ds_3}
    &&\partial^2G_{2\nu\kappa}^{am}(x-y)+gf^{abc}(
		\partial^\mu G_{3\mu\nu\kappa}^{bcm}(0,x-y)+\partial^\mu G_{2\mu\kappa}^{bm}(x-y)G_{1\nu}^{c}(x)
		+\partial^\mu G_{1\mu}^{b}(x)G_{2\nu\kappa}^{cm}(x-y) \nonumber \\
		&&-\partial_\nu G_{3\mu\kappa}^{\mu bcm}(0,x-y)-\partial_\nu G_{2\mu\kappa}^{bm}(x-y)G_{1}^{\mu c}(x)) 
		-\partial_\nu G_{1\mu}^{b}(x)G_{2\kappa}^{\mu cm}(x-y))
		\nonumber \\
		&&+gf^{abc}\partial^\mu G_{3\mu\nu\kappa}^{bcm}(0,x-y)
		+gf^{abc}\partial^\mu(G_{2\mu\kappa}^{bm}(x-y)G_{1\nu}^{c}(x))
				+gf^{abc}\partial^\mu(G_{1\mu}^{b}(x)G_{1\nu\kappa}^{cm}(x-y))
		\nonumber \\
		&&+g^2f^{abc}f^{cde}(G_{4\mu\nu\kappa}^{\mu bdem}(0,0,x-y)
		+G_{3\mu\nu\kappa}^{bdm}(0,x-y)G_{1}^{\mu e}(x) 
		+G_{2\mu\nu}^{bd}(0)G_{2\kappa}^{\mu em}(x-y)\nonumber \\
	&&+G_{3\nu\rho\kappa}^{acm}(0,x-y)G_{1}^{\rho b}(x)
	+G_{2\nu\rho}^{eb}(0)G_{2\kappa}^{\rho dm}(x-y)
	+G_{2\nu\rho}^{de}(0)G_{2\kappa}^{\rho bm}(x-y)
	+G_{1}^{\mu b}(x)G_{3\mu\nu\kappa}^{dem}(0,x-y)+ \nonumber \\
	&&G_{2\kappa}^{\mu bm}(x-y)G_{1\mu}^{d}(x)G_{1\nu}^{e}(x)+
	G_{1}^{\mu b}(x)G_{2\mu\kappa}^{dm}(x-y)G_{1\nu}^{e}(x)+
	G_{1}^{\mu b}(x)G_{1\mu}^{d}(x)G_{2\nu\kappa}^{em}(x-y)) \nonumber \\
	&&	=gf^{abc}(\partial_\nu K^{bcm}_{3\kappa}(0,x-y)+\partial_\nu (\bar P^{b}_1(x)K^{cm}_{2\kappa}(x-y))) 
	+\partial_\nu (\bar K^{bm}_{2\kappa}(x-y)P^{c}_1(x)))
	+ \delta_{am}g_{\nu\kappa}\delta^4(x-y) \nonumber \\
	 &&\partial^2 P^{am}_2(x-y)+gf^{abc}\partial^\mu
	(K^{bcm}_{3\mu}(0,x-y)+P^{bm}_2(x-y)G_{1\mu}^{c}(x)+ \nonumber \\
	&&P^{b}_1(x)K_{2\mu}^{cm}(x-y))=\delta_{am}\delta^4(x-y), \nonumber \\
	&&\partial^2 K^{am}_{2\kappa}(x-y)+gf^{abc}\partial^\mu
	(L^{bcm}_{2\mu\kappa}(0,x-y)+ \nonumber \\
	&&K^{bm}_{2\kappa}(x-y)G_{1\mu}^{c}(x)+P^{b}_1(x)G_{2\mu\kappa}^{cm}(x-y))=0. 
\end{eqnarray}



\end{document}